\documentclass[10pt,twocolumn,a4] {article}
\usepackage{graphicx,colordvi}
\usepackage{amssymb}
\usepackage{xcolor}
\title{\bf Iso-scalar liquid drop model fitting experimental binding energies within macro-micro approach\footnote{Supported by the
 Polish National Science Center (2023/49/B/ST2/01294)  and by the Natural Science Foundation of China  under Grant Nos. 11961131010 and 12335008.}}
 
\author{Krzysztof Pomorski$^{1}$\footnote{Email: pomorski@kft.umcs.lublin.pl},
Zhigang Xiao$^2$\footnote{Email: xiaozg@mail.tsinghua.edu.cn}\\
{\footnotesize$^1$ National Centre for Nuclear Research, Pasteura 7, 02-093 Warsaw, Poland}\\
{\footnotesize $^2$ Department of Physics, Tsinghua University, Beijing 100084,
 China}}

\date{\today}
\begin{document}
\onecolumn

\maketitle

\begin{center} \begin{minipage}{16cm} {\bf Abstract ~}
New liquid drop model with the iso-scalar volume and surface energy terms is applied to reproduce experimentally known masses of nuclei with number of protons and neutrons larger or equal to twenty. The ground-state microscopic energy corrections are taken into account. Even though the model contains only six adjustable parameters {\color{black}in the macroscopic part of the model}, the quality of mass reproduction is good and comparable with other contemporary mass estimates. Also, the fission barrier heights of actinide nuclei evaluated using the topographical theorem of Myers and \'Swi\c{a}tecki are close to the data.
\\



\end{minipage}
\end{center}
\twocolumn
\normalsize

%%%%%%%%%%%%%%%%%%%%%%%%%%%%%%%%%%%%%%%%%%%%%%%%%%%%%%%%%%%%%%%%%%%%%%%%%%%%%%%%
\section{Introduction}

Mass is the most fundamental property of atomic nuclei for its relevance in understanding not only the nuclear structure  and the complex interaction in nuclei, but also  the path of the synthesis of heavy nuclei in astrophysics \cite{LD2003}. In the super heavy region, the mass data also essential in the exploration of synthesizing  super-heavy nuclide in terrestrial laboratory.  Experimentally,  the cooling store ring and  the ion trap both provide  precise measurement of nuclear  mass in modern age \cite{WM2022,ZX2023,YY2024,NIES2025,IRE2025}. So far, approximately  3000-4000 nuclides are measured  while more than 9000 nuclides are believed to exist according to theoretic model calculations \cite{MJ2018} . Descriptions and predictions of nuclear mass are of high theoretic interests and of significance in the calculation of reaction network in stellar environment including r-process \cite{MRM2016}. Many models have been developed for this purpose. Roughly speaking, the models are classified in three groups. i) The macro-micro models, including finite range droplet model (FRDM) \cite{MN95} and Weizsäcker-Skyrme (WS) models \cite{WN2010,LM2011,WN2014},  ii) The pure micro models including Hartree-Fock-Bogoliubov (HFB) models \cite{GS2009}, the covariant density-functional theory (CDFT) \cite{LHZ2015}, {\color{black} relativistic Hartree-Bogoliubov theory \cite{ZCC2024,GCC2025}} etc. and iii) The local mass models which predict the atomic nucleus mass through the relationship of the masses of neighboring nuclides by taking into account different physical constraints \cite{GST2024,YWH2023}, such as Garvey-Kelson (GK) relations \cite{GK1966, MC2020} and the isobaric multiplet mass equation (IMME) \cite{OWE1997} etc. Meanwhile, many efforts have been taken to evaluate and compare the predictive power of the models \cite{NZM2019}. With the applications of the modern machine learning or neural network techniques, the predictive power of the models are further improved  \cite{LYF2021,ZLX2024,LCQ2022,LXK2023,LOVE2022,ZTL2022}.  Up to now, the residues of the nucleus mass reach the level of about 200 keV or better.

%[11] G. T. Garvey, W. J. Gerace, R. L. Jaffe, I. Talmi, and I. Kelson,
%Rev. Mod. Phys. 41, S1 (1969).

Among the macro-micro models,  the liquid-drop (LD) model is one of the oldest nuclear theories. Von Weizsaecker first proposed it \cite{We35} in 1935, and it was generalized in 1936 by Bethe and Bacher \cite{BB36}. This spherical LD model has reproduced with a reasonable accuracy all measured at that time atomic masses. Four years later, Meitner and Frisch \cite{MF39} have added deformation degrees of freedom to the LD model to explain the fission phenomenon discovered by Hahn and Strassmann when bombarding the metallic uranium with neutrons \cite{HS39}. Also in 1939, Bohr and Wheeler proposed this new phenomenon's first theory by expanding the deformed nuclear liquid drop surface in a series of Legendre polynomials \cite{BW39}.

A modern version of the LD model was proposed in 1966 by Myers and \'Swi\c{a}tecki \cite{MS66}. They have shown that the LD energy enriched by the shell and pairing effects can describe well the binding energies and quadrupole moments of known nuclei and gives a reasonable description of the fission-barrier height of heavy nuclei (see also \cite{BD72}). Unfortunately, neither the Myers and \'Swi\c{a}tecki LD formula nor its refined version called the droplet model could adequately reproduce the barrier heights of medium-heavy and lighter nuclei \cite{KN79}. In addition, it was shown by von Groote and Hilf \cite{GH69} that a further correction to the LD model, namely the curvature term, did not change much in this respect. Due to these results further development of the nuclear LD model was stopped in practice for more than three decades. Other much more complicated models like the droplet, Yukawa-folded (YF), Yukawa-plus-Exponential (YpE), or Finite-Range Droplet Models (FRDM) have been used to obtain within the so-called macro-micro approximation \cite{MS66} of the binding energies and the fission barrier heights (for overview look, e.g., Ref.~\cite{KP12}). A more detailed historical review of the LD models can be found, e.g., in Ref.~\cite{Ne09}.

Twenty years ago, it was shown in Ref.~\cite{PD03} that the original model of Myers and \'Swi\c{a}tecki with an additional term containing the curvature energy can simultaneously describe the experimental binding energies of all known at that time isotopes as well as the fission barrier heights. One has to stress that this Lublin-Strasbourg-Drop (LSD) model has reproduced the data with better or comparable accuracy than any other {\it more advanced} theories containing a larger number of adjustable parameters (e.g., Refs.\cite{MN95,MS96}). In the following years, some other parametrizations of the nuclear liquid-drop formula were studied (see, e.g., \cite{Ki08,Ro08,DZ95,MH12,ML12}). Unfortunately, the fission-barrier heights estimated using these models are far from the experimental or estimated ones \cite{Po13}. 

In the present paper, we follow the idea of {\AA}. Bohr and B. Mottelsson \cite{BM69} and Refs. \cite{DZ95,MH12,ML12} to use the quadratic in isospin ($T=|N-Z|/2$) dependence of the nuclear part of the binding energy in the LD formula. We allow here, like in Ref. \cite{MH12}, for a different isospin-square dependence of the volume and surface terms. In addition, the Coulomb exchange energy \cite{Bl29,We35} is taken into account in the present mass formula. Such an extension concerning version (i) of the Moretto et al. LD model \cite{ML12} allows to reproduce much better the experimental masses from the last mass-table \cite{WH21}. Similar to what was done in Refs.~\cite{MS96,PD03,ML12}, we have added to the LD energy the microscopic energy correction evaluated by M\"oller et al. \cite{MS16} when fitting the LD parameters to the experimental masses. One has to stress that the isospin-square LD model of Duflo and Zuker \cite{DZ95,MH12} was later successfully used in several applications (see, e.g., Refs.~\cite{BN14,NB19,Ne19}. That is why we have decided to develop a modern version of the iso-scalar LD model with the parameters adjusted to the atomic masses known at present \cite{WH21}.

%%%%%%%%%%%%%%%%%%%%%%%%%%%%%%%%%%%%%%%%%%%%%%%%%%%%%%%%%%%%%%%%%%%%%%%%%%%%%%%%%%%%

\section{Theoretical model}

A typical nuclear LD formula consists of volume, surface, and Coulomb energy terms:
\begin{equation}
E_{\rm LD}=E_{\rm vol} + E_{\rm sur}\, B_{\rm sur}({\rm def}) 
                       + E_{\rm Coul}\, B_{\rm Coul}({\rm def})~.
\end{equation}
Only the first term is deformation-independent since one assumes that the volume of the nucleus is conserved, while the other terms change with deformation. One has to evaluate their variance assuming some parametrization of the shape of the deformed nuclei. The geometrical, deformation-dependent factors $B_{\rm sur}$ and $B_{\rm Coul}$ have to be evaluated for a given shape parametrization of the deformed nucleus (see, e.g., \cite{BN19}).

We assume the following iso-scalar liquid-drop (ISLD) formula for the energy of a spherical nucleus (see also  Ref.~{\cite{ML12}):
\begin{equation}
\begin{array}{ll}
E_{\rm ISLD}(Z,A;{\rm sph})&=\displaystyle -a_{\rm vol}A\,\left(1-4\kappa_{\rm vol}\frac{T(T+1)}{A^2 }\right)
\\[3ex]
&\displaystyle +a_{\rm sur}A^{2/3}\,\left(1-4\kappa_{\rm sur}\frac{T(T+1)}{A^2 }\right)   \\[3ex]
&\displaystyle +{3\over 5}e^2\frac{Z(Z-1)}{A^{1/3}}+E_{\rm odd}(Z,A) \\[3ex]
&\displaystyle -{3\over 4}e^2\left({3\over 2\pi}\right)^{2/3}\frac{Z^{4/3}}{r_0A^{1/3}}~,
\end{array}
\label{isld}
\end{equation}
were, $Z$ and $A$ are a nucleus's charge and mass numbers, and $e^2=1.43996518\,$MeV$\cdot$fm is the elementary charge square. Here, the volume and the surface part of the binding energies are dependent on the {\color{black} expectation value of the isospin square operator} $T(T+1)$, which is equal to $T=|T_z|= |N-Z|/2$, where $N=A-Z$ is the neutron number. It is easy to show that     
$$4T(T+1)=|N-Z|(|N-Z|+2)$$ 
while in a typical LD model, the term $|N-Z|^2$ is present. 
The odd-even energy $E_{\rm odd}$ is assumed in the following form:
\begin{equation}
E_{\rm odd}(Z,A)=\left\{
\begin{array}{ll}
{\cal D}/Z^{1/3}~& {\rm for~ Z~ odd~,}\\
{\cal D}/N^{1/3}~& {\rm for~ N~ odd~,}\\
{\cal D}/Z^{1/3}+{\cal D}/N^{1/3}~& {\color{black}\rm for~Z~ and~ N~odd~,}\\
 0      & {\rm for~ Z~ and~ N~ even~.}
\end{array}
\right.
\label{odd}
\end{equation}
The last term in Eq. (\ref{isld}) describes the Coulomb exchange energy \cite{We35,Bl29}. 

Note that the linear in isospin term in the ISLD model (\ref{isld}) corresponds to the Wigner (or congruence) energy present in typical LD-like models (refer to, e.g., \cite{MS66,MS96}). In addition, in the ISLD model, the deformation dependence of this linear in $|N-Z|$ is well defined, whereas, in Ref.~\cite{MS96}, one has to add an additional phenomenological deformation dependence in order to obtain a doubling of the congruence at the scission point, when two fission fragment nuclei are born. The only free, i.e., adjustable, parameters of the ISLD model are: $a_{\rm vol},\,\kappa_{\rm vol},\,a_{\rm sur},\,\kappa_{\rm sur},\,r_0$, and ${\cal D}$. 

The following equation gives the mass of an atomic nucleus in the ground state (g.s.):
\begin{equation}
\begin{array}{ll}
M_{\rm th}(Z,A;{\rm g.s.})&= Z\, M_{\rm H}+N\, M_{\rm n}+E_{\rm ISLD}({\rm sph})\\
      &+E_{\rm mic}({\rm g.s.})-\,0.00001433\,Z^{2.39}~,
\end{array}
\label{mass}
\end{equation}
where $M_{\rm H}$=7.289034 MeV and $M_{\rm n}$=8.071431 MeV are the hydrogen and neutron masses measured with respect to the mass unit. The last term approximates the shell energy of electrons. \\
The microscopic energy correction $E_{\rm mic}$ originates from the shell, pairing, and deformation effects:
\begin{equation}
E_{\rm mic}({\rm g.s.})=E_{\rm shell}({\rm g.s.})+E_{\rm pair}({\rm g.s.})
+E_{\rm mac}({\rm g.s.})
\end{equation}
and it is equal to the difference between the ground-state energy of the nucleus and the spherical macroscopic energy. {\color{black} The ground state microscopic energy corrections, take from the tables \cite{MS16}, were evaluated using the FRDM macroscopic deformation energy, the Yukawa-folded single particle potential, and the monopole pairing force.

Using the above microscopic energy correction from the tables \cite{MS16}, we make an approximation. Namely, we assume that the stiffness of the macroscopic deformation energy evaluated within the FRDM is close the that obtained using the ISLD model. A detailed calculation shows that the above approximation may lead to around 50 keV inaccuracy of our mass estimates, which is about ten times smaller than the r.m.s. deviation of the fitted masses shown in Tab. 2. In adition, a similar way of evaluating the nuclear binding energies was also used in Refs. \cite{MS96,PD03,ML12}.}

The Fourier-over-Spheroid (FoS) shape parametrization \cite{Po23,PN23} {\color{black} was used to describe the shape of fissioning nuclei
\begin{equation}
\rho_s^2(z)=\frac{R_0^2}{c}\,f\left(\frac{z-z_{\rm sh}}{z_0}\right)~.
\label{rhos}
\end{equation}
where the function $f(u)$ with $u\in [-1,1]$ is defined as follows:
\begin{equation}
  f(u)=1-u^2-\sum\limits_{k=1}^n \left\{a_{2k}\cos({2k-1\over 2}\pi u)
             +a_{2k+1}\sin(k\pi u)\right\}
\label{fos}
\end{equation}
Here $\rho_s(z)$ is the distance of a surface point to the $z$-axis and $z_0=c R_0$, with $R_0$ being the sphere's radius, is the half-length of the deformed nucleus. The first two terms in $f(u)$ describe a spheroid, while the others give the deviation of the nucleus surface from the spheroidal form. The shift parameter $z_{\rm sh}= -3/(4\pi)\,z_0(a_3-a_5/2+ \dots)$ ensures that the origin of the coordinate system is located at the center of mass of the nucleus. The volume conservation condition implies $ a_2=a_4/3-a_6/5+\dots$. The expansion coefficients $a_i$ are treated as the deformation parameters. The parameter $c$ determines the elongation of the nucleus, keeping its volume fixed, while $a_3$ and $a_4$ are, respectively, the deformation parameters essentially responsible for the reflection asymmetry and the neck formation of the deformed shape.

The FoS parametrization reproduces} nuclear shapes very close to the {\it optimal shapes} obtained by the Strutinsky variational procedure \cite{IP09} and allows to evaluate precisely the liquid drop energy at the saddle point. Knowing this energy, one can estimate the fission barrier height ($V_{\rm sadd}$) with the help of the Myers and \'Swi\c{a}tecki topographical theorem \cite{MS96}
\begin{equation}
V_{\rm sadd}= M_{\rm mac}({\rm sadd}) - M_{\rm exp}({\rm g.s.}) \,\,,
\label{TT}
\end{equation}
where $M_{\rm mac}({\rm sadd})$ is the macroscopic mass at the saddle point and $M_{\rm exp}({\rm g.s.})$ is the ground-state experimental mass. The argument of \'Swi\c{a}tecki in favor of Eq. (\ref{TT}) was that the shell, or better to say microscopic energy corrections at the saddle-point, are small as the fissioning system, tries to avoid hills and dales on its way to fission. Of course, such argumentation is only valid when one discusses the energy of the highest saddle point, not the deformation of the nucleus at the saddle point. It was shown in Ref.~\cite{DN08} that the above rough approximation reproduces fairly well the experimental fission barrier heights. 

%%%%%%%%%%%%%%%%%%%%%%%%%%%%%%%%%%%%%%%%%%%%%%%%%%%%%%%%%%%%%%%%%%%%%%%%%%%%%%%%%%%%%%%%%%%%%%%%%%%%%%%%%%%%%%%%%%%%%%%%
\section{Nuclear masses in different macro-micro models}

In the last issue of the atomics mass table \cite{WH21}, there are 2259 measured and 906 estimated masses of isotopes with $Z,N\ge 20$ having an experimental error smaller than 1.5 MeV. All these masses are taken to find the best set of six adjustable parameters {\color{black} in the marcroscopic part} of our ISLD model (\ref{isld}). The mean square deviation of theoretical estimates and the experimental masses is minimized to obtain the best set of parameters. The microscopic energy $E_{\rm micr}$ from the M\"oller et al. mass-table \cite{MS16} were used when evaluating the masses of isotopes using Eq.~(\ref{mass}). Two parameter sets are found: one (a) corresponding to 2259 measured masses and the second one (b) obtained using 3165 measured and estimated masses denoted by the hash sign in the mass-table \cite{WH21}. The r.m.s. deviation, which is a measure of the fit quality, is taken in the following form:
\begin{equation}
\sigma= {1\over n}\sum\limits_{i=1}^n (M_{\rm th}-M_{\rm exp})^2~,
\label{sigma}
\end{equation}
where $i$ runs over all isotopes taken into account. We have also evaluated $\sigma$ for three traditional models: Thomas-Fermi (TF) of Myers and \'Swi\c{a}tecki \cite{MS96}, LSD of Pomorski and Dudek \cite{PD03}, FRDM of M\"oller et al. \cite{MS16}. The results are presented in Table I.

\begin{table*}[t!]
\caption{Root mean square deviations (in MeV) of the experimental masses \cite{WH21} and the theoretical ones evaluated in different models.}
\begin{center}
\begin{tabular}{|c|c|c|c|c|c|}
\hline &&&&&\\[-2ex]
Number of nuclei  &   TF    &   LSD   &  FRDM   & ISLD(a) & ISLD(b)  \\[0.5ex]
%              &        &         &         &  (a) &  (b)  \\[0.5ex]
\hline &&&&&\\[-2ex]
 Exp.~~~~~~~ 2259 \hfill  &  0.669  &  0.523  &  0.536  &{\bf 0.517}& 0.620 \\[0.5ex]
\hline &&&&&\\[-2ex]
 Exp.+est. 3165  &  0.874  &  0.817  &  0.956  &  0.917    &{\bf 0.745}\\[0.5ex]
\hline\hline &&&&&\\[-2ex]
No. adj. param. & 8 + 4  & 8 + 4   &   9 + 4   &     6     &      6    \\[0.5ex]
\hline 
\end{tabular}
\end{center}
\end{table*}

\begin{table}[h!]
\caption{The parameter set of the ISLD model (Eq. \ref{isld}) fitted to the experimental (a) and experimental plus estimated (b) atomic masses.}
\begin{center}
\begin{tabular}{|c|c|c|c|}
\hline &&&\\[-2ex]
   Parameter      & Units &  ISLD (a)  & ISLD (b)   \\[0.5ex]
\hline\hline &&&\\[-2ex]
$a_{\rm vol}$     &  MeV  & -15.381912 & -15.499610 \\[0.5ex]
\hline &&&\\[-2ex]
$\kappa_{\rm vol}$&   -   &   1.87161  &   1.88029  \\[0.5ex]
\hline &&&\\[-2ex]
$a_{\rm sur}$     &  MeV  &  17.27931  &  17.64004  \\[0.5ex]
\hline &&&\\[-2ex]
$\kappa_{\rm sur}$&   -   &   2.3022   &   2.2727   \\[0.5ex]
\hline &&&\\[-2ex]
$r_0$             &  fm   &   1.229286 &   1.214608 \\[0.5ex]
\hline &&&\\[-2ex]
${\cal D}$        &  MeV  &  4.305     &  4.234     \\[0.5ex]
\hline
\end{tabular}
\end{center}
\end{table}

It is seen that the ISLD model with only six adjustable parameters fitted to the experimental data for 2259 isotopes (a) reproduces the isotope masses with even better quality than the FRDM from which the microscopic energy corrections are taken. Of course, the r.m.s. deviation of the theoretical and measured masses grows when one makes the fit to all 3165 experimental and estimated masses. So, an additional fit (b) of the ISLD parameters was performed when all 3165 isotope masses were considered. Both sets (a) and (b) of the ISLD parameters are listed in Table II. Surprisingly, the 28-year-old Thomas-Fermi model predicts these 906 additional masses better than the FRDM. Also, the LSD model developed in 2003 describes both experimental data and all experimental and estimated data very well, proving its good predictive power. The last row in Table I shows the number of adjustable parameters of each model. The Thomas-Fermi model \cite{MS96} has 8 free parameters plus four additional parameters for the congruence (Wigner) and odd-even energy taken Ref.~\cite{MN95}. Also, the Lublin-Strasbourg Drop model is based on the congruence and odd-even energy (4 parameters) developed in Ref.~\cite{MN95} and has eight adjustable constants. The FRDM has nine plus four fitted parameters corresponding the the macroscopic part of energy, while the present ISML model is based on six adjustable parameters only.
\begin{figure}[t!]
\includegraphics[width=\columnwidth]{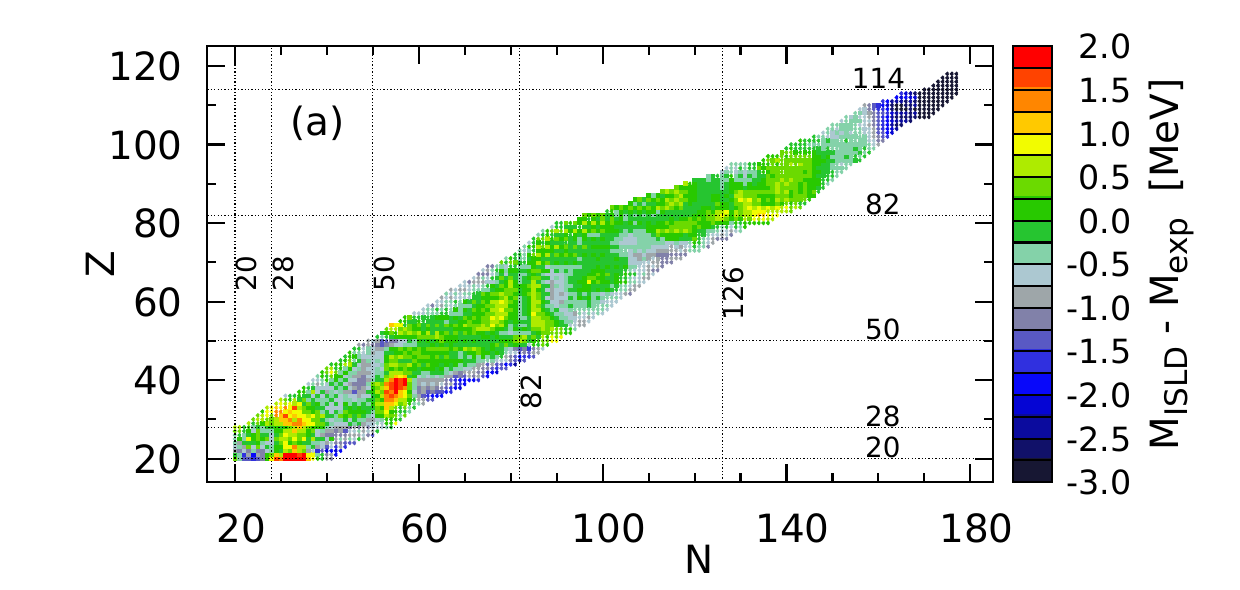}\\
\includegraphics[width=\columnwidth]{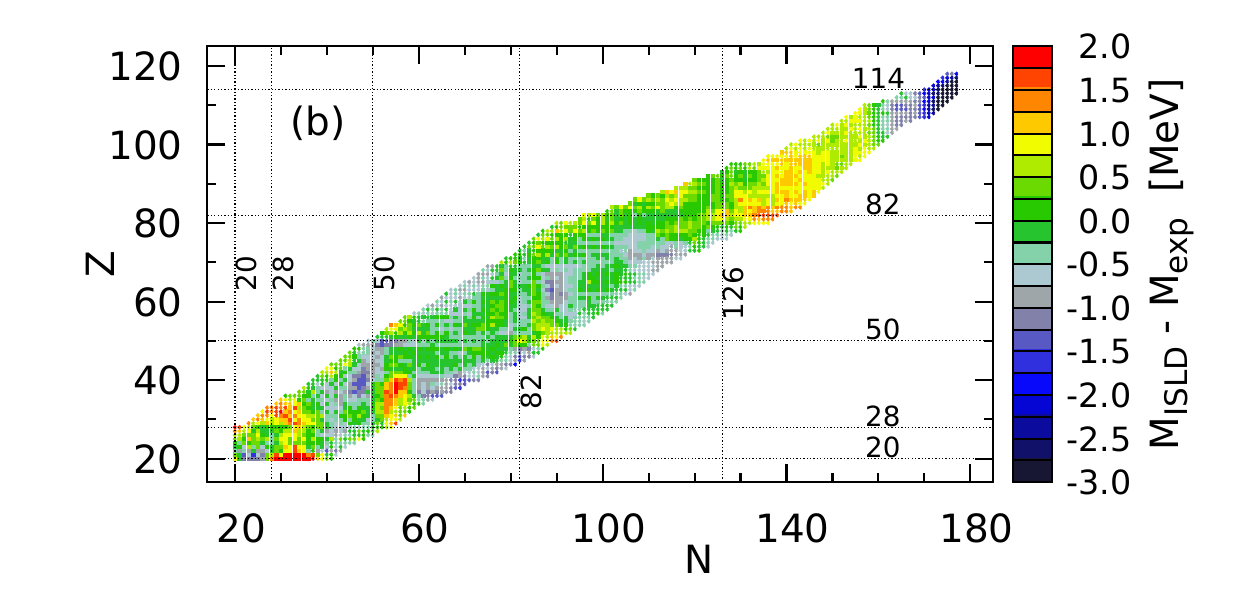}
\caption{Deviation of the atomic mass estimates obtained using the ISLD(a) (top) and ISLD(b) (bottom) from experimental data (squares). Circles mark the deviations from the estimated data.}
\label{MISLD-Mexp}
\end{figure}

The deviations between the estimated and the experimental (squares) and the estimates (circles) masses are shown in Fig.~\ref{MISLD-Mexp} for the ISLD parameter set (a) (top) and (b) (bottom). It is seen that significant discrepancies appear for neutron-rich or proton-rich nuclei and superheavy nuclei, where the majority of masses are estimated (circles). In addition, some discrepancies originating probably from poorly reproduced shell effects are visible. The largest ones are in the vicinity of Z=20 and 28 magic numbers and around $^{94}$Zr. 

\begin{figure}[t!]
\includegraphics[width=\columnwidth]{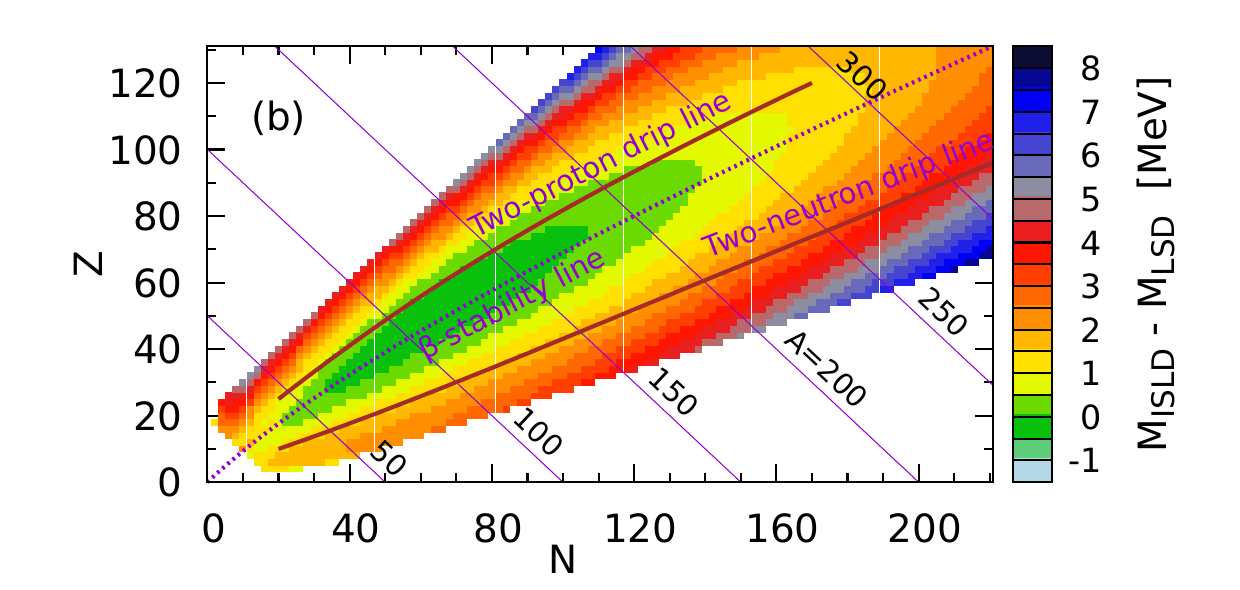}
\caption{Difference of the atomic mass estimates obtained using the {\color{black} ISLD(b) and the LSD \cite{PD03} models.}}
\label{MLSD-MISLD}
\end{figure}
A question appears, which set of the ISLD parameters should be used: the one fitted to the experimental masses only (a) or adjusted to all experimental and estimated masses (b)? One has to note that more than half of the isotopic masses of heavy nuclei with mass-number $A\ge 220$ listed in the mass table~\cite{WH21} corresponds to the estimated, not the measured data. In addition, the pure experimental masses are only less than 10\%  of the superheavy nuclei (SHN) data with $Z\ge 104$. So, the set (b) of the ISLD parameters is recommended when describing the properties of the heaviest nuclei. In the following section, we shall use and refer to the set (b), and we will call it simply the ISLD set of parameters.

\begin{figure*}{h!}
\includegraphics[width=\columnwidth]{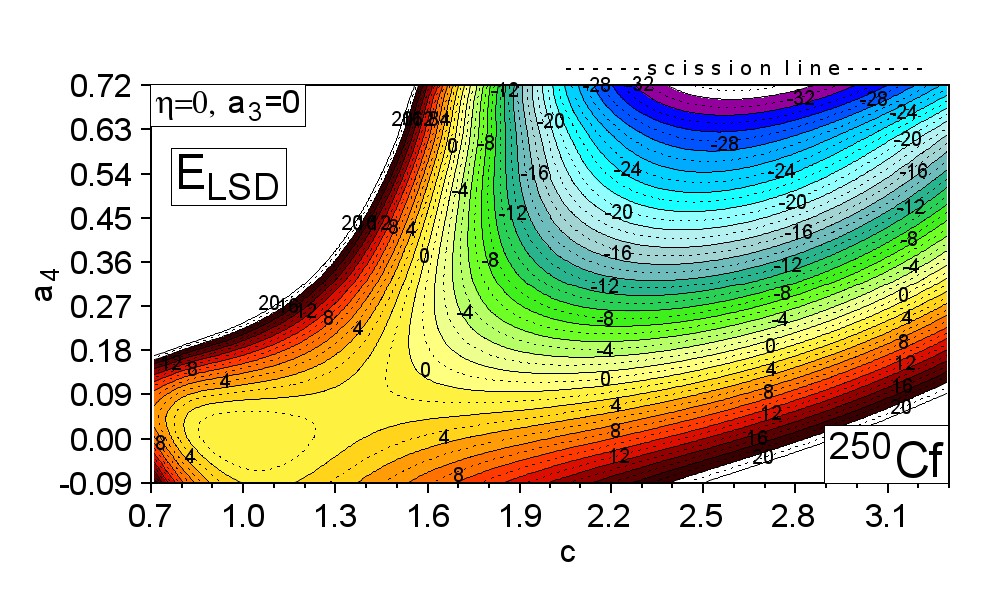}
\includegraphics[width=\columnwidth]{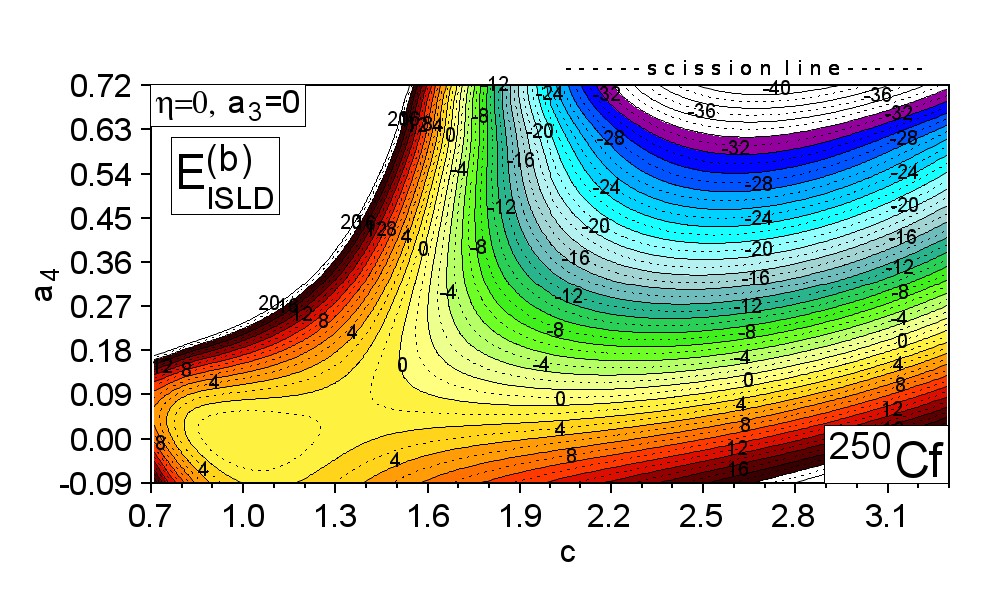}\\[-2.5ex]
\includegraphics[width=\columnwidth]{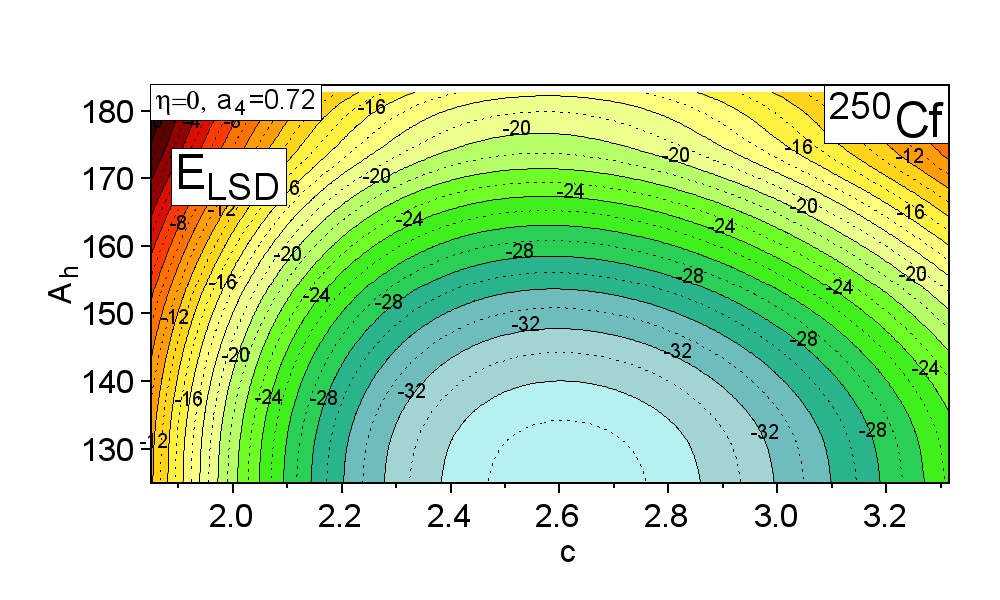}
\includegraphics[width=\columnwidth]{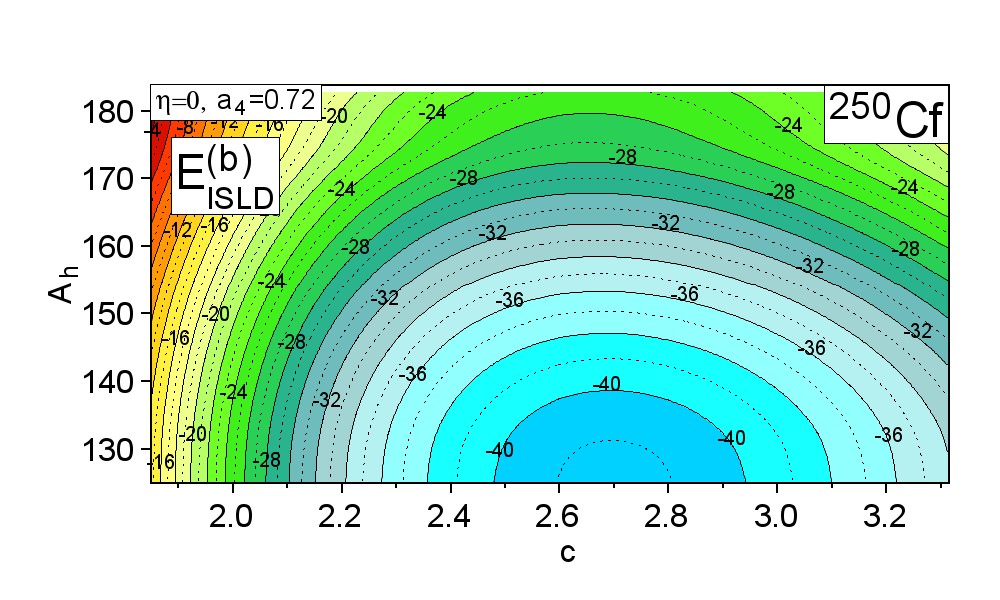}\\[-4ex]
\caption{Macroscopic energy surfaces of $^{250}$Cf on the ($c,a_4$) plane (top) and around the scission configuration (bottom) are evaluated using the LSD (l.h.s) c and ISLD(b) (r.h.s.) formulae. Here $c$ is the elongation of the nucleus, $a_4$ is related to the neck size, and ${\rm A_h}$ is the heavy fission fragment mass number \cite{PN23}.}
\label{epot}
\end{figure*}

One has to mention here that we have also performed mass fits by using a similar as the LD formula (\ref{isld}) but with the curvature term proportional to $A^{1/3}$ and with the Coulomb redistribution energy ($\sim Z^2/A$). However, adding such two terms to the ISLD formula did not change significantly the r.m.s. deviation from the data, so, finally, they are not considered in our model.
 
It is well known that the nuclear masses predictions for nuclei close to the $\beta$-stability line obtained within different macroscopic models are close to each other. Significant differences between the models may appear when one goes toward the proton or neutron drip lines. This effect is illustrated in Fig.~\ref{MLSD-MISLD}, where the differences between the ISLD(b) and LSD mass estimates are shown as a function of neutron ($N$) and proton ($Z$) numbers. It is seen that the differences do not exceed the range (-0.5,+0.5) MeV for isotopes with $A \le 220$ laying between the two-proton drip and $\beta$-stability lines. Also, both estimates are close to each other for neutron reach isotopes close to the $\beta$-stability line. In the region of superheavy nuclei and for isotopes close to the two-neutron drip line, the ISLD masses are even 1.5 MeV larger than the LSD ones. Such differences in the mass estimates may be significant for predicting the astrophysical r-process or the stability and decays of the SHN.

%%%%%%%%%%%%%%%%%%%%%%%%%%%%%%%%%%%%%%%%%%%%%%%%%%%%%%%%%%%%%%%%%%%%%%%%%%%%%%%%%%%%%

\section{Potential energy surfaces and fission barrier heights}

The LSD and ISLD macroscopic energy surfaces of $^{250}$Cf are compared in Fig.~\ref{epot}. We have used here the FoS shape parametrization \cite{PN23}. It was also assumed that the macroscopic energy of a spherical nucleus ($c=1,\,a_4=0$) is zero. It is seen that the surfaces obtained in both models are close to each other, and only tiny differences can be noticed. Namely, the ISLD fission valley (top row) is slightly deeper and corresponds to more elongated shapes than the LSD one. The exit of the LSD valley around the scission configuration (bottom raw) is located at the elongation $c=2.62$. It is lying around 4 MeV above the ISLD exit, which appears at $c=2.70$. Such differences in energy and elongation of the nucleus in the macroscopic scission point may have some influence on the total kinetic energy (TKE) of the fission fragments. On the other hand, the stiffnesses of LSD and ISLD potentials with respect to the fission fragment mass asymmetry ($A_h$) (see the maps in the bottom raw) are very close to each other. The saddle points in both ($c,\,a_4$) maps correspond to almost the same energy ($E_{\rm sadd}\approx 1.3$ MeV and are located around the deformation $c\approx 1.35$ and $a_4\approx 0.06$. 

\begin{figure}
\includegraphics[width=\columnwidth]{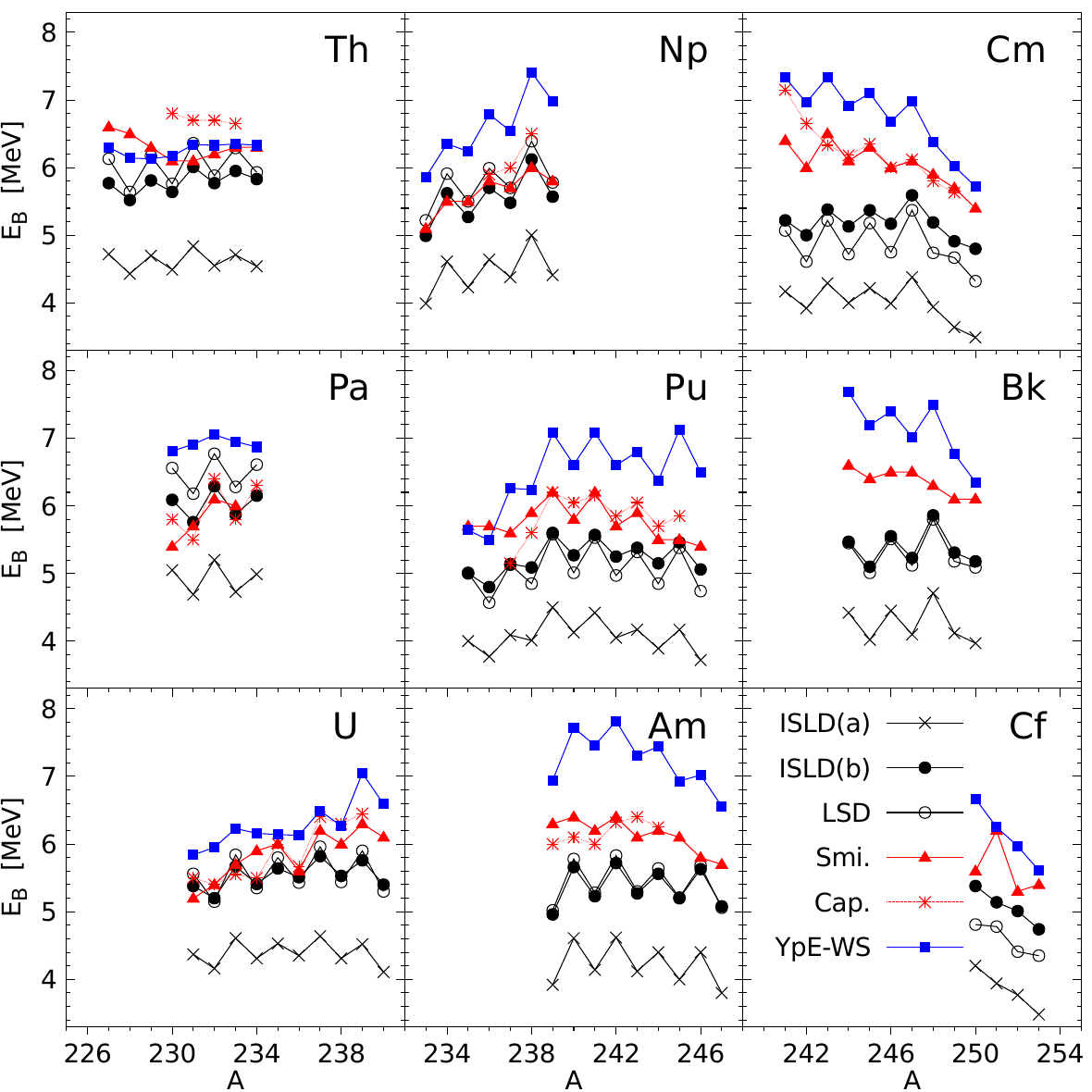} 
\caption{Fission barrier heights evaluated using the topographical theorem {\color{black}(see Eq.~\ref{TT})} with the ISLD(a) (crosses), ISLD(b) (points), and LSD (open circles) set of parameters are compared with the empirical/experimental data taken from Ref.~\cite{Sm93} (triangles) and \cite{Ca09} (stars) and with the theoretical values (squares) obtained within the macro-micro model in Ref.~\cite{JK21}.}
\label{ebh}
\end{figure}

Such macroscopic saddle-point energies will be used in the following to estimate the fission barrier heights of the actinide nuclei by the topographical theorem of Myers and \'Swi\c{a}tecki (\ref{TT}).

The fission barrier heights estimated using Eq.~(\ref{TT}) and the ISLD (a) (crosses) and (b) (circles) and the LSD (open circles) formulas are compared in Fig.~\ref{ebh} with the experimental data taken from Refs.~\cite{Sm93} (triangles) and \cite{Ca09} (stars). In addition, the theoretical estimates of the highest fission barrier done in Ref.~\cite{JK21} are marked by squares. The experimental, or better to say {\it empirical}, fission barrier heights come mainly from analysis of the fission cross-section energy dependence and fissionability of nuclei. They are supplemented by data obtained from analysis of the excitation functions of spontaneously fissioning isomers and the group of strong resonances in the sub-barrier fission cross-section \cite{Sm93,Ca09}. The theoretical barrier heights tabulated in Ref.~\cite{JK21} were evaluated within the 7D macro-micro model with the YpE macroscopic energy part and the microscopic energy evaluated using the Woods-Saxon (WS) single-particle potential. 

As one can see, the fission barrier heights obtained using the Myers and \'Swi\c{a}tecki topographical theorem and the LSD and ISLD models underestimate, in most cases, the experimental barrier heights while those of Ref.~\cite{JK21} are, as a rule, larger than the experimental values. The ISLD barriers obtained with the set (b) parameters are very close to the LSD ones, while those computed with the set (a) are, on average, by 1 MeV smaller. One can expect that the fission barriers evaluated in a similar as in Ref.~\cite{JK21} way, but with the ISLD macroscopic part of the energy will be closer to the data as the topographical theorem estimates the barrier heights from below \cite{MS96}.

\section{Conclusions}

We have shown that the new liquid drop mass formula with the charge asymmetry term proportional to the isospin square (ISLD) describes the presently known experimental and estimated from systematic atomic masses well. One mast stress that the ISLD formula {\color{black} for the macroscopic energy} contains only six adjustable parameters. The other models contain more free parameters, e.g., the twenty-one years old LSD mass formula \cite{PD03}, which reproduces the binding energies even better than the FRDM theory \cite{MS16}, possesses eight directly fitted parameters and four others (in the congruence/Wigner and odd-even terms) taken from the adjustments made in Ref.~\cite{MN95}. In this place, it is good to remember that microscopic models, typically having up to 14 interaction parameters, can reproduce 2149 experimental masses with the r.m.s. deviation equal to 0.798 MeV \cite{GH09}.

It was found that all experimentally estimated masses of superheavy nuclei with a neutron number larger than 160 have up to 3 MeV larger masses than the predicted ones. What is the origin of these discrepancies: inaccuracies in evaluating the microscopic energy, wrong asymptotic behavior of our macroscopic models, or inaccurate evaluation of the atomic masses from the systematics? More detailed calculations to answer this question are necessary.
  
It was also shown that both ISLD (b) and LSD models describe well the fission barrier heights of the heavy nuclei. The ISLD model predicts slightly larger atomic masses than the LSD one in the superheavy region of nuclei and for neutron-rich isotopes close to the neutron drip line. This difference in the mass estimates could be significant for SHN physics, and it can influence the nuclear r-process probabilities, which is very important in astrophysical theories. Also, the ISLD model predicts a more elongated shape of fissioning nuclei at scission configuration than the LSD formula. This effect can influence estimates of the fission fragment TKE and their charge equilibration (refer to Ref.~\cite{PN24}). 

We plan to perform extended dynamical calculations like those made in Refs.~\cite{PN23,PN24,PB21,KD21,LC21} but the use of the new ISLD formula when evaluating the potential energy surfaces of fissioning nuclei. The well-defined deformation dependence of the macroscopic energy term linear in $\vert N-Z\vert$ may also influence the isotopic yields of the fission fragments (refer to Ref.~\cite{PN24}). \\

\noindent
{\bf Acknowledgements}\\[+1ex]
The author thanks Kai Neerg{\aa}rd for his helpful remarks.

\end{document}